\documentclass[aps, prl, twocolumn, showpacs, floatfix, superscriptaddress, groupedaddress]{revtex4-1}
\usepackage[utf8]{inputenc}
\usepackage{amsthm}
\usepackage{amssymb}
\usepackage{amsmath}
\usepackage{gensymb}
\usepackage{graphicx}
\usepackage{url}
\usepackage{framed}
\usepackage{float}
\usepackage{wrapfig}
\usepackage{multirow} 
\usepackage{tikz}
\usepackage{fullpage}
\usepackage[toc,page]{appendix}
\usepackage{physics}
\usepackage{listings}
\usepackage{hyperref}
\hypersetup{
    colorlinks=true,
    linkcolor=blue,
    citecolor=blue,
    filecolor=magenta,      
    urlcolor=cyan,
}

\usepackage{geometry}
\usepackage{natbib}

\usetikzlibrary{arrows}
\usetikzlibrary{decorations.markings}

\usepackage{csquotes}
\MakeOuterQuote{"}

\begin{document}


\title{Contacts and upstream modes explain the electron-hole asymmetry in the graphene quantum Hall regime}

\author{N. Moreau$^1$, B. Brun$^{1}$, S. Somanchi$^2$, K. Watanabe$^3$, T. Taniguchi$^4$, C. Stampfer$^2$ \& B. Hackens$^1$}

\affiliation{
$^1$ IMCN/NAPS, Universit\'e catholique de Louvain (UCLouvain), B-1348 Louvain-la-Neuve, Belgium \\
$^2$ JARA-FIT and 2nd Institute of Physics - RWTH Aachen, Germany \\
$^3$ Research Center for Functional Materials, National Institute for Materials Science, Namiki, Japan \\
$^4$ International Center for Materials Nanoarchitectonics, National Institute for Materials Science, Namiki, Japan
}

%

\date{\today}

\begin{abstract}
Observations of electron-hole asymmetry in transport through graphene devices at high magnetic field challenge prevalent models of the graphene quantum Hall effect. Here, we study this asymmetry both in conventional magnetotransport and in scanning gate microscopy maps measured in an encapsulated graphene constriction. We reveal that the presence of upstream modes and local doping in the vicinity of electrical contacts leads to a totally different picture of topological breakdown for electrons and holes, explaining the observed asymmetry. 
\end{abstract}

\maketitle



Injecting and collecting charges in two-dimensional electronic systems (2DESs) through ohmic contacts look like easy tasks, with most models describing contacts as smooth extensions of the 2DES. However, in the quantum Hall (QH) regime, the topic is far from trivial as charge carriers flow in different topologically-protected quantum Hall edge channels (QHECs). Indeed, good coupling of these channels with the contacts involve no backscattering of incoming QHECs and perfect equilibration in the injection of carriers in the different outgoing QHECs. Fulfilling these two criteria proves difficult because of the complex interface between the metallic contacts and the 2DES \cite{Komiyama1989, Dahlem2010, Weis2011}.

In the graphene QH regime, the situation is even more cumbersome, for two reasons at least. First, state-of-the-art QH devices rely on graphene encapsulation in hexagonal boron nitride (hBN) and on one-dimensional line-contacts obtained by depositing metal on the flanks of the device \cite{Wang2013}. Such abrupt metal-graphene contact was shown to potentially destroy electron-hole symmetry \cite{Khomyakov2010}. Second, different experimental results indicate that upstream (i.e. counterpropagating) QHECs can flow along device edges \cite{Cui2016, Marguerite2019, Moreau2020}. The most common explanation ascribes their coexistence to an inhomogeneous screening of the back gate potential, related to fringing fields, leading to charge carriers accumulation at device borders \cite{Silvestrov2008}. 

Thanks to scanning gate microscopy (SGM), topological breakdown of the QH effect has recently been shown to originate from the coupling of up- and downstream QHECs along the very same edge in this material \cite{Marguerite2019, Moreau2020}. In particular, we showed that the coupling is achieved through the localized states of antidots located along the edges \cite{Moreau2020}. Importantly, the latter SGM experiments have essentially focused on hole-type charge carriers and evidences for the same mechanisms for electron-type charge carriers remain elusive. 


In this letter, we combine transport and SGM with tight-binding simulations to study the topological breakdown of graphene QHECs in the case of electrons and holes. Our results reveal that contacts lead to a different spatial configuration of upstream QHECs for both types of charge carriers. This asymmetry is at the origin of distinct mechanisms causing a topological breakdown on the electron and hole sides.

\begin{figure}[!ht]
\centering
\includegraphics[width=\linewidth]{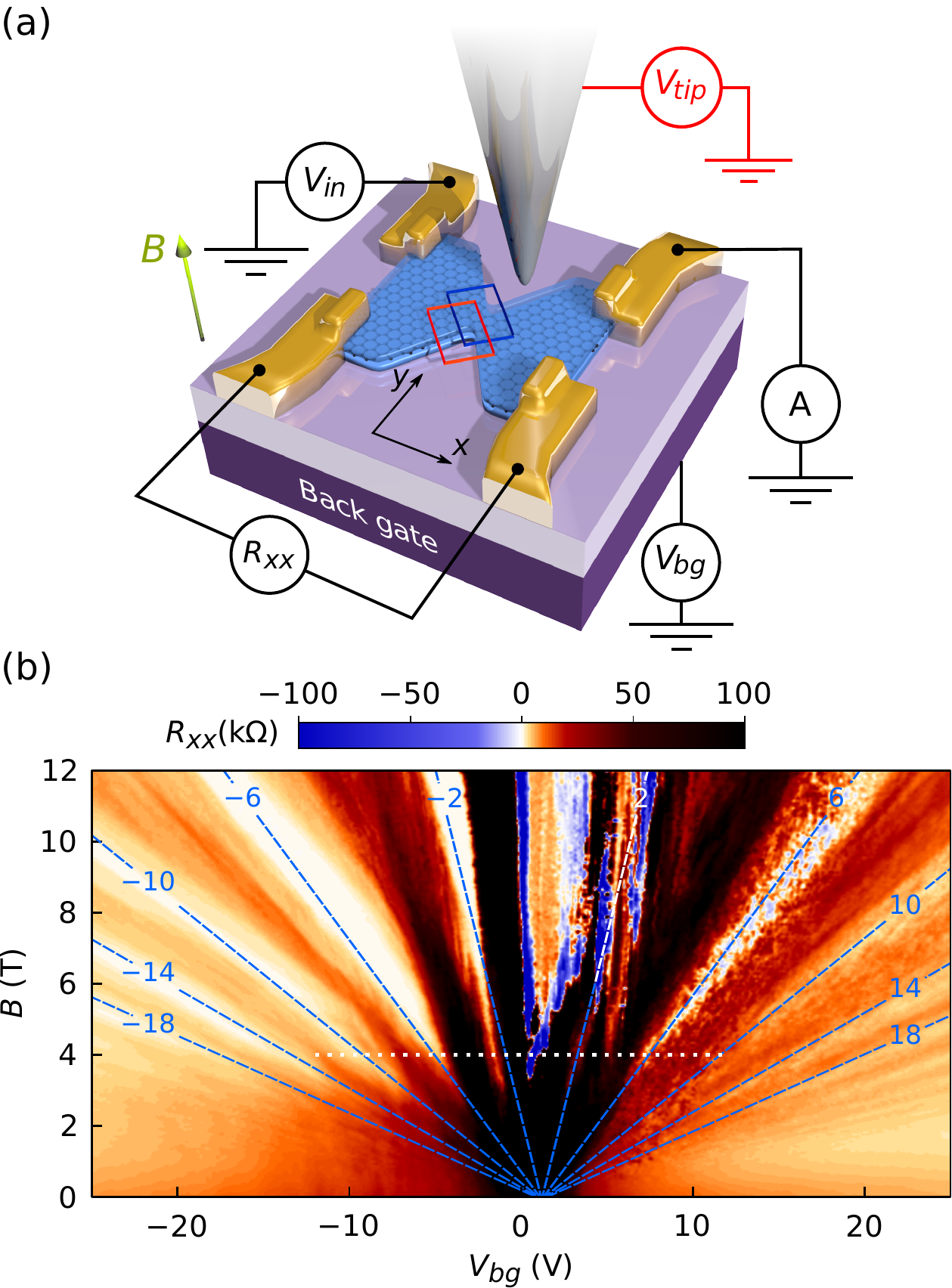}
\caption{(a) Artist view of the experimental setup. (b) $R_{xx}$ map as a function of $V_{bg}$ and $B$. The blue lines indicate the positions of the filling factors $\nu = \pm (4n+2)$.}
\label{sample}
\end{figure}

The studied sample, depicted in Fig. \ref{sample}a, consists in a monolayer of graphene encapsulated between two hBN flakes \cite{Terres2016}. The longitudinal resistance $R_{xx}$ is measured via line contacts made of gold deposited on a thin adhesion layer of chromium \cite{Wang2013}. Inherently to this contact layout, they overlap the heterostructure. Charge carrier density is varied using a back gate voltage $V_{bg}$ and a magnetic field $B$ is applied perpendicularly to the graphene plane. We furthermore use a sharp metallic SGM tip, biased at a voltage $V_{tip}$ to change locally the charge carrier density. 

\begin{figure}[!ht]
\centering
\includegraphics[width=\linewidth]{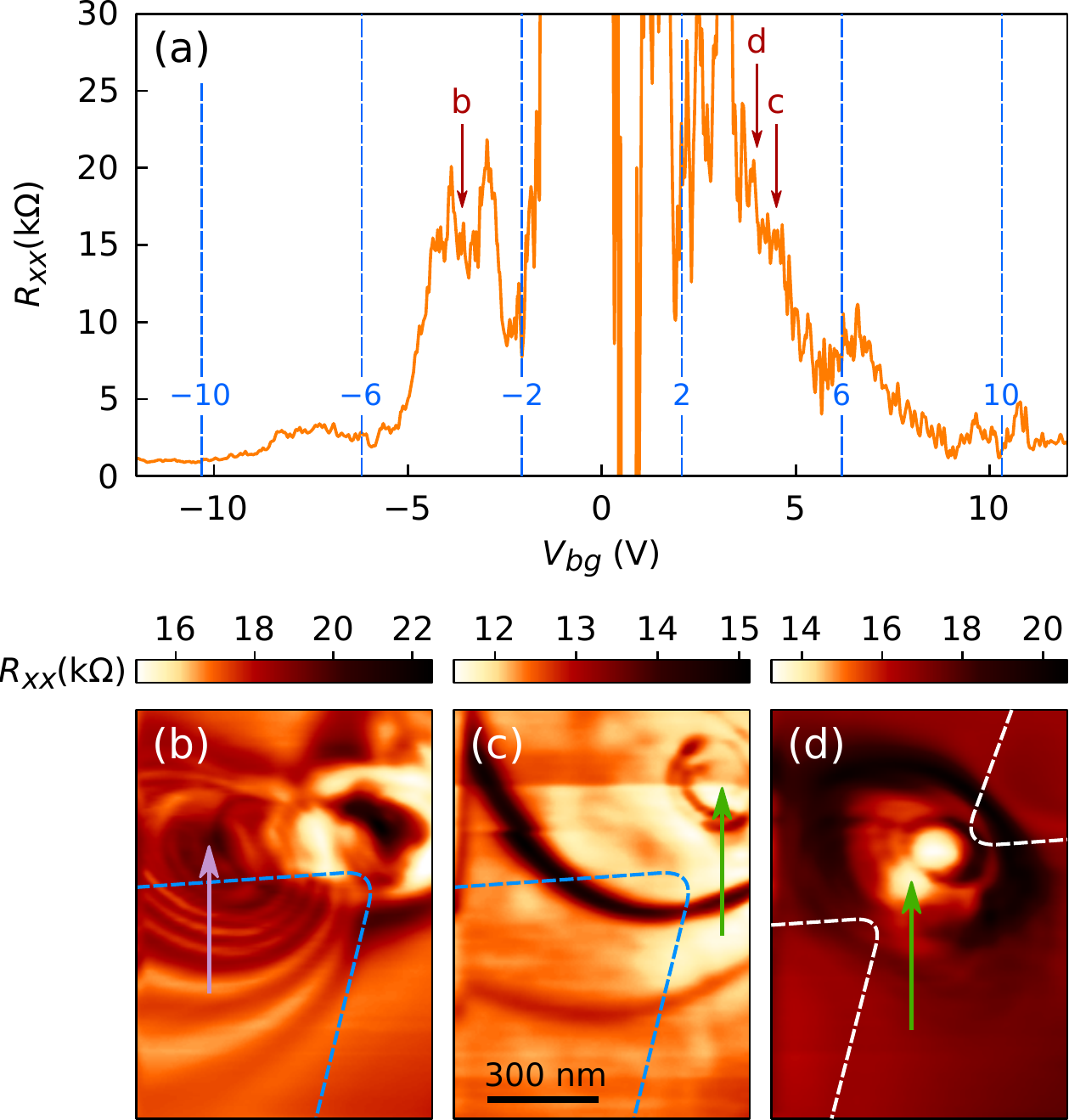}
\caption{(a) $R_{xx}$ as a funcion of $V_{bg}$ at $B = 4$ T (white dotted line in Fig. \ref{sample}b). The blue dashed lines indicate the positions of the filling factors $\nu = \pm (4n+2)$. (b-d) SGM maps obtained in the vicinity of the constriction for holes (b) and electrons (c,d). The scan area is depicted in Fig. \ref{sample}a with a red rectangle for (b,c) and with a dark blue one for (d). The $V_{bg}$ values are indicated with arrows in (a) and $V_{tip} = 0.5$ V in (b,c) and $1.5$ V in (d). For holes, SGM contrast appears centered on the sample edge (purple arrow) as well as in the center of the constriction whereas it is located exclusively at the center of the constriction for electrons (green arrows).}
\label{SGM}
\end{figure}

A fan diagram is presented in Fig. \ref{sample}b. The white stripes correspond to vanishing $R_{xx}$ and indicate that QHECs are topologically protected, so that charge carriers backscattering is forbidden. At low positive $V_{bg}$ and above $B = 4$ T, $R_{xx}$ exhibits extremely large and irregular fluctuations. Such an asymmetric and perturbed behavior has already been reported in \cite{Amet2015} and is relatively common in the case of encapsulated graphene devices. In order to stand at the verge of this regime, we limit our measurements to $B=4$ T in the remainder of this paper. $R_{xx}$ as a function of $V_{bg}$ is shown in Fig. \ref{SGM}a for this value of magnetic field.

SGM maps, obtained by recording $R_{xx}$ as a function of the tip position, reveal precious information on the origin of the coupling between QHECs. The tip perturbation locally tunes the coupling and, hence, the backscattering between QHECs \cite{Hackens2010, Martins2013, Marguerite2019, Moreau2020, Paradiso2011, Pascher2014}. In turn, this leads to concentric fringes of $R_{xx}$ (local maxima or minima) in the SGM map, centered on the spot where the coupling occurs. As shown in Fig. \ref{SGM}b, SGM contrasts is located along edges for holes. It is coherent with former studies that ascribed these signatures to the presence of upstream QHECs, coupled with the downstream modes flowing along the same edge \cite{Moreau2020, Marguerite2019}. For electrons, however, SGM contrast is centered only in the constriction region, as indicated with SGM fringes with higher $R_{xx}$ in Fig. \ref{SGM}c and with spots of lower $R_{xx}$ in Fig. \ref{SGM}d. Contrary to the hole case, no contrast is found centered on the edges of the device. It is noteworthy that similar signatures have been found in SGM experiments on classical semiconductor-based 2DESs \cite{Hackens2010, Martins2013}, where there is no upstream QHEC. In these systems, topological breakdown was assigned to the presence of an antidot located in the vicinity of the constriction and coupling the QHECs running at the opposite device edges.

As a first hypothesis to account for electron-hole asymmetry, one could speculate that upstream QHECs only exist for holes. Impurities and defects along the borders could in this case favor holes accumulation and induce such charge asymmetry, as proposed by Marguerite et al. \cite{Marguerite2019} (referred to hereafter as the "impurity model"). While our data can not undoubtedly discard this scenario, the impurity model alone fails to yield a full picture of QHECs in graphene in regards to other experimental results reported in the literature \cite{Cui2016} (see supplemental materials). On the other hand the mainstream theory, namely the inhomogeneous screening of the backgate potential (the fringing field model), predicts the emergence of upstream QHECs both for holes and electrons, and consequently an apparent charge symmetry. In the following, we show that electron-hole asymmetry can also be explained in the fringing field model framework.

In the model that we develop hereafter, we assume that charge carriers accumulate along the edges in the same way on the electron and hole sides. The bending of energy bands leads in both cases to the presence of upstream QHECs. In this scenario, electron-hole asymmetry originates from the contacts. In particular, the difference of work-function between graphene and chromium, forming the adhesion layer of contacts, yields a hole-type doping below the metallic regions overlapping the stack illustrated in Fig. \ref{schematics}a. The work-function-induced difference of the potential (see supplemental materials) can be approximated by \cite{Giovannetti2008, Chaves2014}
\begin{equation}\label{CrGr}
    \Delta \phi = \dfrac{\sqrt{1+2\alpha e|W_{\text{Cr}}-W_{\text{G}}|}-1}{\alpha},
\end{equation}
where $W_{\text{Cr}} = 4.5$ eV \cite{Waldrop1984} and $W_{\text{G}} = 4.48$ eV \cite{Giovannetti2008} are the work-functions of chromium and graphene respectively and $\alpha = 2e^2t_{\text{hBN}}/(\epsilon_{\text{hBN}}\pi \hbar^2 v_F^2)$ with $t_{\text{hBN}}\sim 20$ nm the vertical distance between graphene and the overlapping metal, $2<\epsilon_{\text{hBN}}<4$ the hBN permittivity and $v_F\sim 10^6$ m/s the Fermi velocity of graphene. With these data, we find $0.006 < \Delta \phi < 0.008$ eV. As a result, charge carrier density is larger below the overlapping metal, with respect to the bulk, when the bulk is doped with holes, whereas it is smaller for an electron-doped bulk.

\begin{figure}[!ht]
\centering
\includegraphics[width=\linewidth]{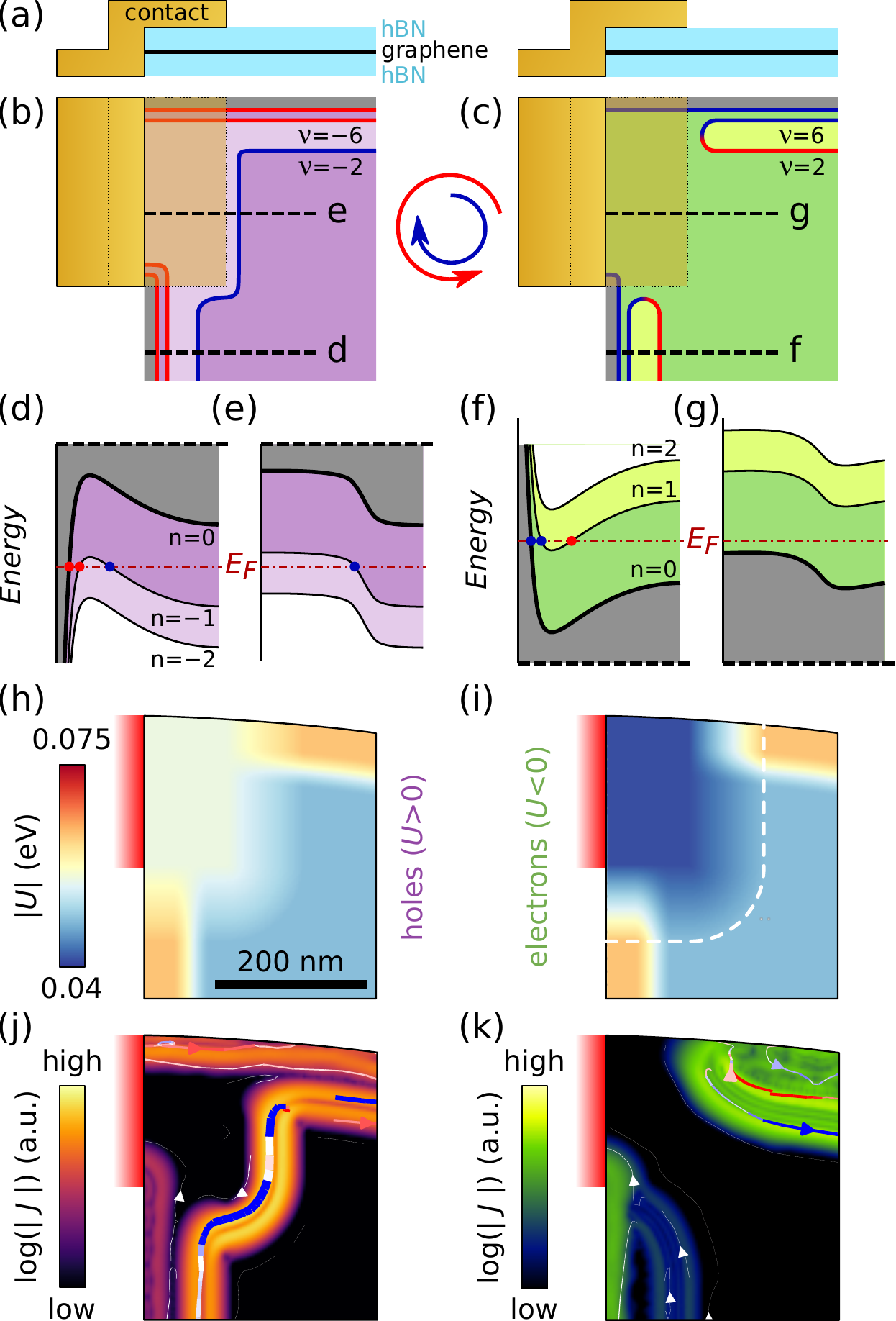}
\caption{(a) Side view of the device (schematic) in the vicinity of one of the contacts. (b,c) Top view of one of the sample corner, around the contact. Due to the increase of charge carrier density at the device borders, upstream QHECs run along the edges (clockwise in blue and counterclockwise in red). If the contacts dope graphene with holes, the inner (clockwise) QHEC for holes circumvent the contact area (b) whereas the inner (counterclockwise) QHEC for electrons is merged with the nearest external (clockwise) QHEC (c). (d-g) Evolution of the potential (thick line) and the three first Landau levels (LLs) along the black dashed lines in (b) and (c). QHECs appear where LLs cross the Fermi energy (red dashed line). (h,i) Absolute value of the onsite potential landscape $|U|$ used for the tight-binding simulations for holes with $V_{bg} = -2.4$ V (h) and electrons with $V_{bg} = 2.4$ V (i). The semi-infinite leads in the simulated system are depicted in red. (j,k) Current density maps obtained from tight-binding calculations for holes (j) and electrons (k).}
\label{schematics}
\end{figure}

In the QH regime, this modulation of charge carrier density has an important influence on the QHECs spatial configuration in the vicinity of the contacts, which differs for holes and electrons, as illustrated in Figs. \ref{schematics}b,c. In this schematic, upstream QHECs flow along the edges where one of the bent Landau levels crosses twice the Fermi energy, as shown in Fig. \ref{schematics}d for holes and \ref{schematics}f for electrons. However, this picture is no longer valid around and below the contacts pads, i.e., in the doped orange-shaded region in Fig. \ref{schematics}b-c. For holes, the inner (clockwise) QHEC circumvent the contact zone, due to the increase of holes density (Fig. \ref{schematics}e). For electrons, the lower charge density close to the contacts (Fig. \ref{schematics}g) leads to the merging of the inner (counterclockwise) with the closest clockwise QHECs (Fig. \ref{schematics}c). These QHECs are therefore perfectly equilibrated.

Based on this model, we performed tight-binding simulations, using the KWANT package \cite{Groth2014}. Maps of the absolute value of the on-site potential around the contacts are shown in Fig. \ref{schematics}h for holes and in Fig. \ref{schematics}i for electrons. They feature an increase of charge carrier density along the edges, proportional to the bulk density (see supplemental materials), in addition to a constant positive offset energy $\Delta E = 0.007$ eV (Eq. (\ref{CrGr})) around the contacts. The resulting current density maps are presented in Figs. \ref{schematics}j,k. The QHECs are visible in lighter tones and match the qualitative picture of Figs. \ref{schematics}b,c.

\begin{figure}[!ht]
\centering
\includegraphics[width=\linewidth]{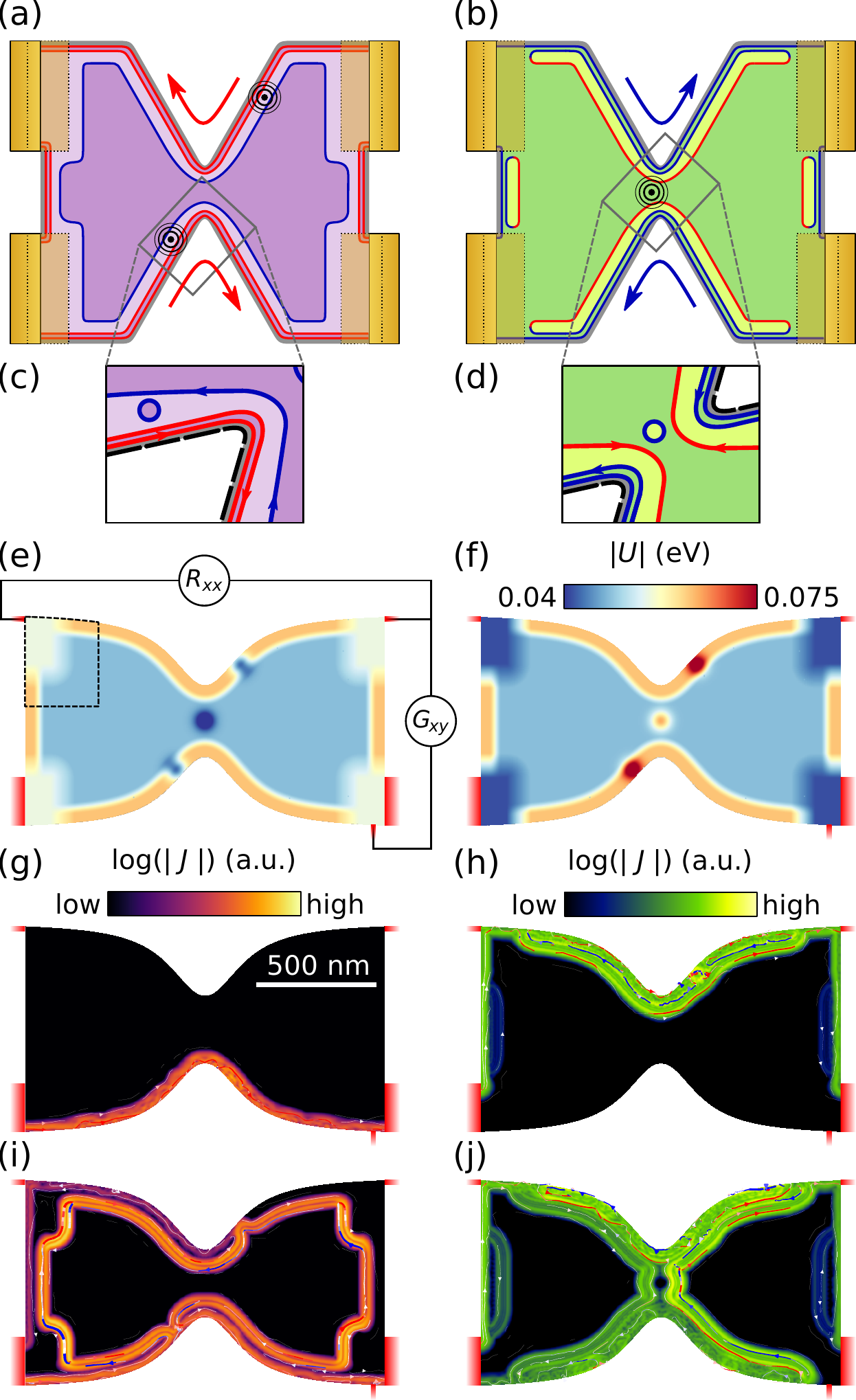}
\caption{(a) For holes, the inner QHEC describes a closed loop so that the effective filling factor is given by the light purple area’s order (here $\nu = -6$). (b) For electrons, the inner (counterclockwise) QHEC forms a closed loop as it connects with the nearest clockwise QHEC so that the filling factor is given by the dark green area’s order (here $\nu = 2$). (c) For holes, SGM signatures along the edges in Fig. \ref{SGM}b are caused by the coupling between upstream QHECs occurs via an antidot located close to the constriction. (d) For electrons, coupling between opposite border QHECs via an antidot. (e,f) Potential landscapes of the simulated systems, with colors corresponding to the absolute value of the onsite potential landscape $|U|$, with the same parameters as in Figs. \ref{schematics}h,i. (g-j) Current density maps obtained for holes with $V_{bg} = -4.8$ V (g), $-3.8$ V (i) and electrons with $V_{bg} = 6.4$ V (h), $7.8$ V (j) as indicated with arrows in Fig. \ref{Rxx}a.}\label{simus}
\end{figure}

We now consider the entire sample to examine the consequences of these different QHECs spatial configurations at the contacts for holes and electrons, and the relationship with the experimental data of Fig. \ref{SGM}. For holes (left side of Fig. \ref{simus}), it appears that the inner QHEC (blue) forms a closed loop, as illustrated in Fig. \ref{simus}a. In the absence of coupling with the outer QHECs (red), this loop is topologically equivalent to a localized state in QH theory since it is not directly coupled with contacts and should not influence transport. The effective filling factor of the system is therefore given by the region outside the loop (here $\nu = -6$, in light purple). To observe the topological breakdown of QHECs, holes running along one of the sample edges should be backscattered to the opposite edge. This can only be achieved by coupling the upstream QHEC loop both on the upper and lower sides of the sample, for example at locations pinpointed by concentric black circles in Fig. \ref{simus}a. The SGM contrast highlighted along the sample edge in Fig. \ref{SGM}b is therefore the signature of one of the two coupling spots between upstream QHECs. In particular, the coupling is achieved through an antidot, as depicted in Fig. \ref{simus}c \cite{Moreau2020}. By introducing these antidots in the simulated potential landscape, with Gaussian functions centered close to the edges (see the potential map in Fig. \ref{simus}e), we correctly capture the mechanism leading to holes backscattering. Fig. \ref{simus}g shows a simulated current density map where propagating QHECs are not coupled through the antidots. This is in stark contrast with Fig. \ref{simus}i where coupling is achieved at a slightly different value of $V_{bg}$ (calculated from $E_F$), so that holes flow through the inner QHEC loop, connecting both edges.

For electrons (right side of Fig. \ref{simus}), the equilibrated QHECs form loops running along the edges, as illustrated in Fig. \ref{simus}b. Contrary to holes, the effective filling factor of the system is given by the bulk (here $\nu = 2$, in dark green), as for conventional semiconductor-based 2DEGs. The topological breakdown of QHECs can only be achieved by coupling the upstream QHECs of the upper and lower sides of the sample, which can only happen in the vicinity of the constriction, as illustrated with concentric black circles in Fig. \ref{simus}b. Electrons running along one of the edges (in a QHEC loop) can then be backscattered to the opposite side of the device. The SGM contrast highlighted at the center of the constriction in Figs. \ref{SGM}c-d is therefore the consequence of the coupling between the opposite QHECs loops through an antidot located at the center of the constriction, as depicted in Fig. \ref{simus}d \cite{Hackens2010}. Simulations support this picture, when introducing an antidot at the center of the constriction (Fig. \ref{simus}f). Fig. \ref{simus}h depicts a current density map where QHECs loops are visible along the edges but the absence of coupling with the opposite edge prevents backscattering. When coupling is active (at different $V_{bg}$, as in Fig. \ref{simus}j), electrons flow from one edge to the other through the central antidot and can be backscattered.

\begin{figure}[!ht]
\centering
\includegraphics[width=\linewidth]{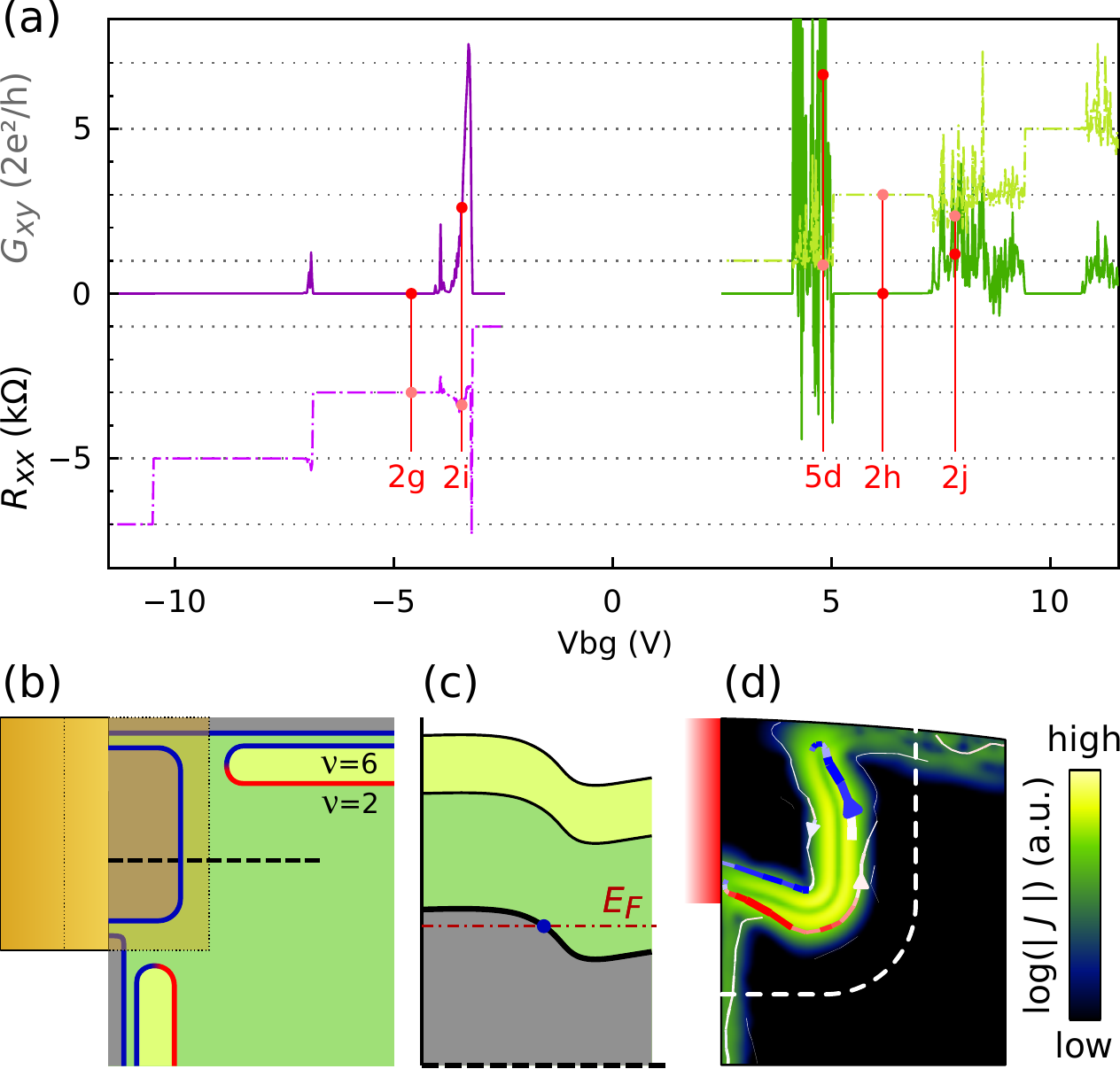}
\caption{(a) Simulated $R_{xx}$ (plain line) and $G_{xy}$ (dotted line) as a function of $V_{bg}$ for a magnetic field $B = 4$ T. (b) Spatial configuration of the QHECs at the origin of the fluctuations in (a). (c) Spatial evolution of the potential (thick line) and of the three first Landau levels (LLs) along the black dashed lines in (b). (d) Current density maps obtained from tight-binding calculations for $V_{bg} = 4.8$ V, as indicated with a red line in (a).}\label{Rxx}
\end{figure}

The simulated $R_{xx}$ and $G_{xy}$ curves as a function of $V_{bg}$, obtained in this system, are presented in Fig. \ref{Rxx}a for holes (purple) and electrons (green). When $R_{xx}$ is zero, $G_{xy}$ exhibits the expected plateaus for graphene at $4(n + 1/2)e^2/h$. Non-zero $R_{xx}$, accompanied by the transition between two Hall plateaus, indicates the coupling between the opposite device edges (as shown in Fig. \ref{simus}) and the non-zero probability for charge carriers to be back-scattered. A striking feature of the curve is the presence of large $R_{xx}$ fluctuations on the electron side, similarly to the experimental data presented in Fig. \ref{SGM}a. These fluctuations originate from extra QHECs running below the contacts, inducing a direct backscattering of charge carriers in the contact area (Figs. \ref{Rxx}b,c). This mechanism prevents a proper equilibration between the propagating QHECs and the contacts. In addition, current injection is strongly hampered in this situation, preventing proper determination of $R_{xx}$. Parasitic doping below the contacts therefore explains measurements artifacts such as the large variation of resistance even towards negative values reported here (Fig. \ref{SGM}a) as well as in previous experiments \cite{Amet2015}.

In summary, we have highlighted that line contacts, used for state-of-the-art encapsulated graphene samples, can lead to different spatial configurations of the upstream QHECs on the electron and hole sides. This yields distinct mechanisms at the origin of the QH topological breakdown for both charge carriers types. For holes, the breakdown occurs by coupling the upstream QHECs through antidots located at the device edges whereas for electrons, all the QHECs running along the same edge are equilibrated and the breakdown occurs through the bulk, similar to conventional semi-conductor 2DEGs. This interpretation is in full agreement with microwave impedance microscopy results, where the QH topological breakdown was shown to coincide either with a conducting bulk, either with conducting edges, depending on the charge carriers type \cite{Cui2016}. Our conclusions pinpoint the importance of considering the influence of contacts when designing two-dimensional materials-based samples, in particular when these contacts are to be coupled with topologically protected edge channels.

\begin{acknowledgments}
The authors thank Samuel Dechamps and Jean-Christophe Charlier for discussion about metal-graphene interaction. This work was partly funded by the Federation Wallonie-Bruxelles through the ARC Grant No. 16/21-077, by the F.R.S-FNRS through the Grant No. J008019F, and from the European Union’s Horizon 2020 Research and Innovation program (Core 1 No. 696656 and Core 2 No. 785219). This work was also partly supported by the FLAG-ERA grant TATTOOS, through F.R.S.-FNRS PINT-MULTI grant No. R 8010.19. B.B. (research assistant), B.H. (research associate) and N.M. (FRIA fellowship) acknowledge financial support from the F.R.S.-FNRS of Belgium. Support by the Helmholtz Nanoelectronic Facility (HNF), the EU ITN SPINOGRAPH and the DFG (SPP-1459) is gratefully acknowledged. K.W. and T.T. acknowledge support from the Elemental Strategy Initiative conducted by the MEXT,
Japan ,Grant Number JPMXP0112101001, JSPS KAKENHI Grant Numbers JP20H00354 and the CREST(JPMJCR15F3), JST.

\end{acknowledgments}

%



\end{document}


\title{\textbf{\LARGE{Contacts and upstream modes explain electron-hole asymmetry in graphene quantum Hall regime \\ - Supplemental materials -}}}

\author{\large{N. Moreau$^1$, B. Brun$^{1}$, S. Somanchi$^2$, K. Watanabe$^3$, }\\ \large{T. Taniguchi$^4$, C. Stampfer$^2$ \& B. Hackens$^1$} \\
\small{$^1$IMCN/NAPS, Universit\'e catholique de Louvain (UCLouvain), B-1348 Louvain-la-Neuve, Belgium} \\
\small{$^2$ JARA-FIT and 2nd Institute of Physics - RWTH Aachen, Germany} \\
\small{$^3$ Research Center for Functional Materials,}\\
\small{National Institute for Materials Science, 1-1 Namiki, Japan} \\
\small{$^4$ International Center for Materials Nanoarchitectonics,}\\
\small{National Institute for Materials Science, 1-1 Namiki, Japan}}


\date{}

\maketitle

\section{Experimental setup}
The studied sample consists in a monolayer of graphene, encapsulated between two 20 nm-thick hBN flakes, in which a 250 nm-wide constriction geometry has been lithographically-defined \cite{Terres2016}, as depicted in Fig. 1a of the manuscript. It is anchored to the mixing chamber of a dilution fridge whose base temperature is $\sim 100~\text{mK}$. A magnetic field $B$ is applied perpendicularly to the graphene plane. Current is injected and voltage is measured via line contacts made of gold deposited on a thin adhesion layer of chromium \cite{Wang2013}. Inherently to this contact layout, they overlap the heterostructure over an area of $\sim1\times 0.2~ \mu \text{m}^2$, as pictured in Fig. 1a of the manuscript. The four contacts allow to measure the longitudinal resistance $R_{xx}$ using the standard lock-in technique at $77~\text{Hz}$. The global charge carrier density in graphene is varied by applying a voltage $V_{bg}$ on the silicon back-gate separated from the sample by a 300 nm-thick SiO$_2$ layer. We furthermore use a sharp metallic SGM tip, biased at a voltage $V_{tip}$, and scanning at a distance of 70 nm from the graphene plane, to change locally the charge carrier density. 

\section{Quantum Hall regime at low magnetic field}
\begin{figure}[!ht]
\centering
\includegraphics[width=.6\linewidth]{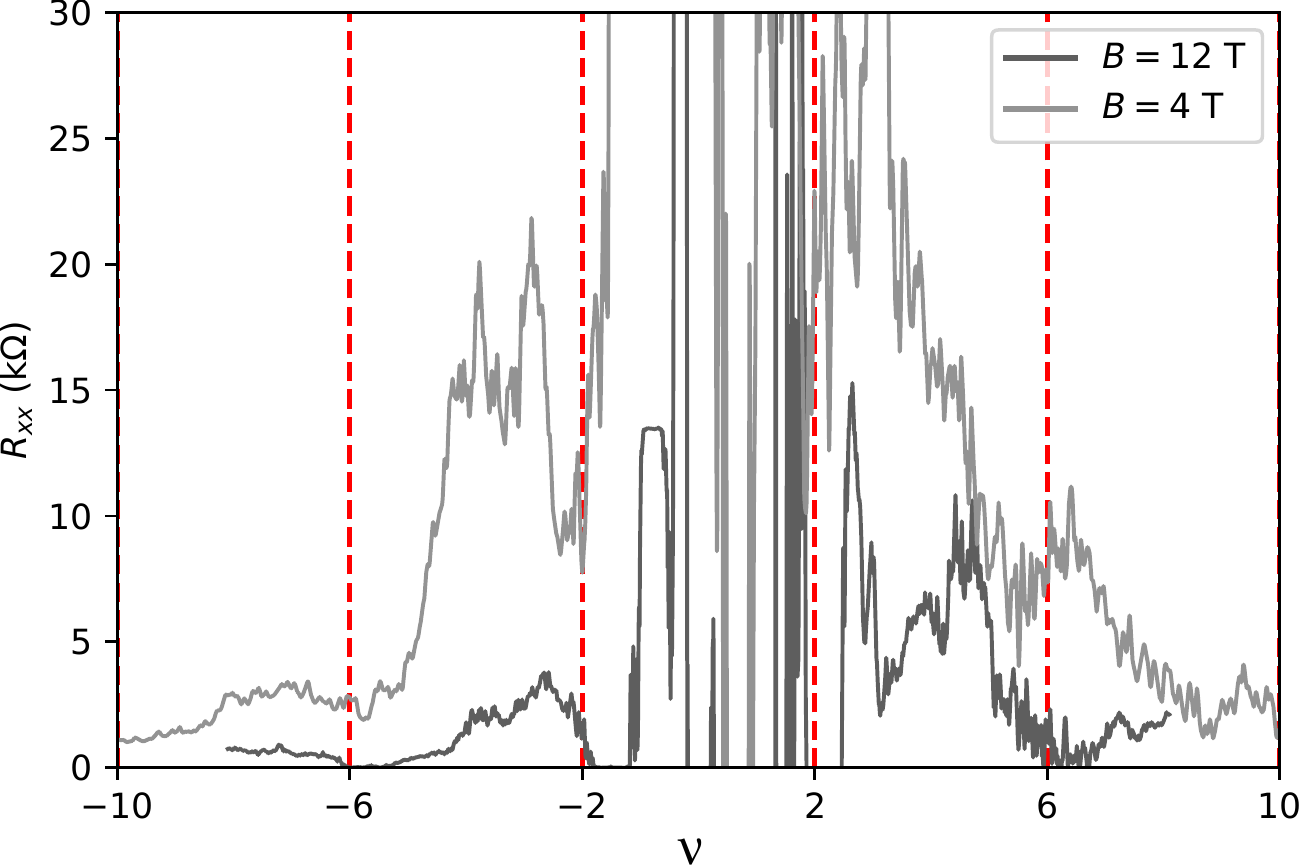}
\caption{Longitudinal resistance $R_{xx}$ as a function of the filling factor $\nu$ for a magnetic field of 4 T (light gray) and 12 T (dark gray) and a temperature of 100 mK. The values of $\nu = \pm 4(n+1/2)$ at integer $n$ are indicated with red dashed lines.}
\label{Rxxnu}
\end{figure}

In Fig. 2a of the manuscript and Fig. \ref{Rxxnu}, we present the longitudinal resistance $R_{xx}$ as a function of the back gate voltage $V_{bg}$ (converted in filling factor $\nu$ in Fig. \ref{Rxxnu}), recorded at a magnetic field $B=4$ T. In this data at relatively low $B$, the topological protection of quantum Hall channels (QHECs) appears to be poor, since $R_{xx}$ exhibits minima at the filling factors $\nu = \pm 4(n+1/2)$ but does not fall to zero as expected in the quantum Hall (QH) regime. It must be compared to measurements performed at higher magnetic field (see the Fan diagram in Fig. 1b of the manuscript and the curve for $B = 12$ T in Fig. \ref{Rxxnu}), where $R_{xx}$ falls to zero, or is close to zero, around $\nu = \pm 4 (n+1/2)$, indicating a stronger topological protection of QHECs. 

\begin{figure}[!ht]
\centering
\includegraphics[width=.9\linewidth]{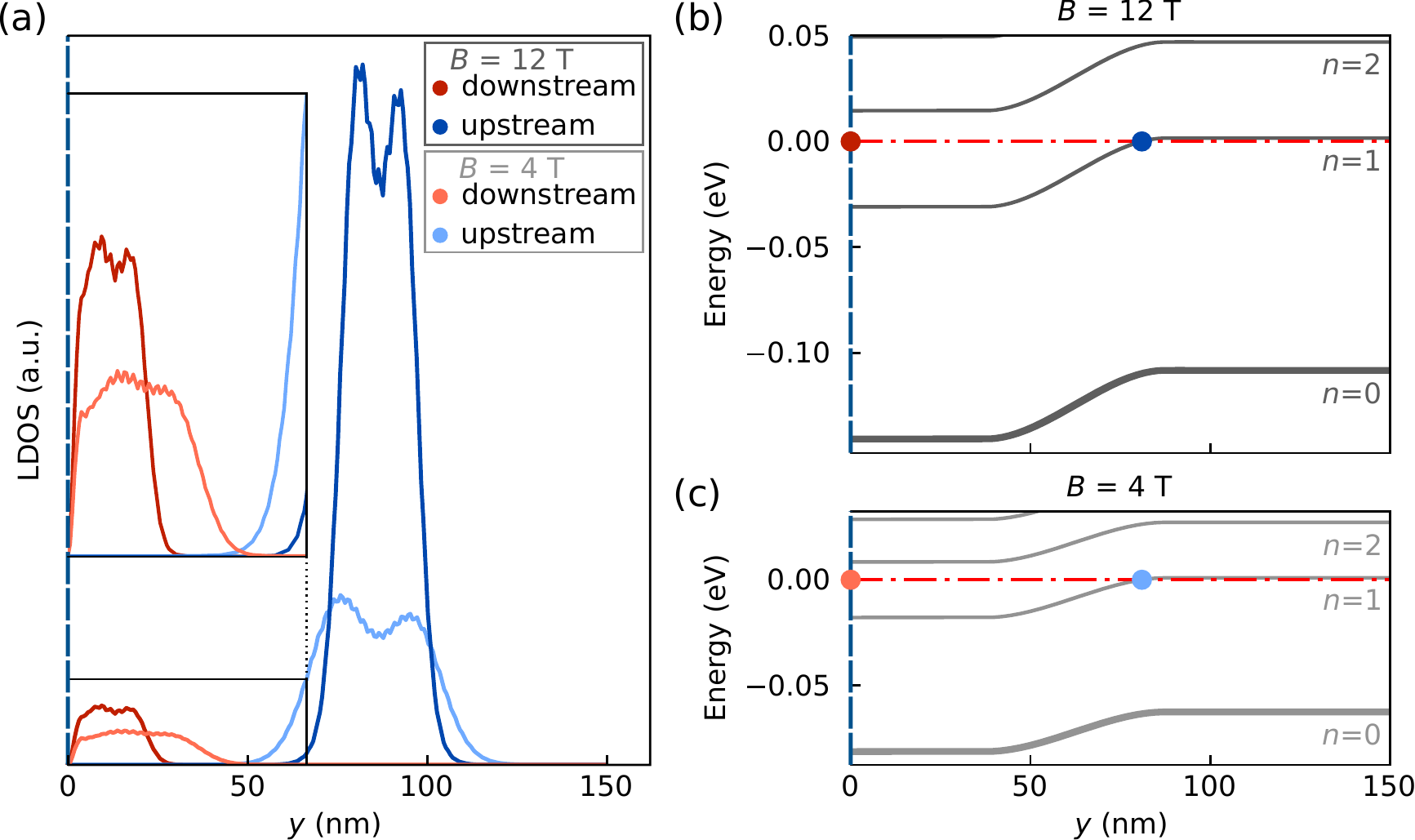}
\caption{Width of up- and downstream QHECs at $B=4$ and 12 T. \textbf{(a)} KWANT simulation of the LDOS as a function of the distance $y$ from the graphene edge. The LDOS associated to downstream (upstream) QHECs is presented in light red (blue) for $B = 4$ T and in dark red (blue) for $B = 12$ T. \textbf{(b,c)} Potential as a function of the distance $y$ from the graphene edge (thick line) used to simulate the LDOS of (a) and the three first Landau levels ($n=0,1,2$) for $B = 12$ T (b) and $B = 4$ T (c). The potential has been qualitatively designed to induce an increase of the charge carriers density at the edges proportional to the bulk density by a factor $\eta = 0.69$ (a factor 0.3 between edge and bulk onsite parameters, as detailed in section \ref{suppl_simus}). Upstream modes (blue dots) arise when the Fermi energy (red dashed-dotted line) crosses one of the bent Landau level. Downstream modes (orange dots) appear due the infinitely sharp confinement potential at the graphene edge.}
\label{LDOS_up}
\end{figure}

To understand the poor topological protection of QHECs at $B = 4$ T, one must note that the width of the QHECs decreases when the magnetic field increases. Indeed, the typical width of the wavefunction associated to a QHEC is given by the magnetic length $l_B = \sqrt{\hbar/eB}$ \cite{Goerbig2011}. Since both up- and downstream QHECs flow along the same edge in graphene, they can overlap at low magnetic field while they are kept away from each other at higher field, as illustrated in Fig. \ref{LDOS_up}a. However, an overlap of the counterpropagating QHECs does not mean they are coupled. The equilibration of these channels requires the presence of impurities between them, allowing a transfer of charge carriers from down- to upstream modes \cite{Paradiso2011}. These impurities are at the origin of the antidots revealed with the SGM measurements presented in the manuscript, whose associated localized QHECs also have a varying overlap with propagating QHECs. To conclude, the absence of zero resistance states at 4 T around $\nu = \pm 4 (n+1/2)$ in Fig. \ref{Rxxnu} can be explained by the proximity of the counterpropagating QHECs combined with the presence of impurities located between these channels.

\section{More about the impurity model}\label{imp_model}
In the manuscript, we evoke a first hypothesis to explain the difference of behavior between holes and electrons observed in Figs. 2b-d. In that framework, impurities accumulated along the edges induce a hole doping at the borders of the sample. Considering this mechanism alone, upstream QHECs would therefore only exist on the hole side. We have named this scenario "the impurity model". Signatures of such impurities, affecting the structure of QHECs, have been reported in two recent papers \cite{Marguerite2019,Cui2016}. Here, we summarize their findings and we discuss why the upstream QHECs can not be explained solely in the impurity model framework. We furthermore shed a new light on those results by analyzing them in the scope of our model, presented in the manuscript.

Recently, Marguerite \textit{et al.} \cite{Marguerite2019} evoked the possible presence of impurities doping the graphene edges with holes. They came to this conclusion on the basis of ingenious transport measurements, both in the electrons and hole sides, while varying the voltage of top plunger gates intended to locally change the charge carriers density at the sample edges (Extended Data Figs. 5 and 7 in \cite{Marguerite2019}). Coupled to SGM measurements, their data revealed the signature of a holes accumulation along the edges while the sample's bulk was doped with electrons. As a consequence, the increase of hole density at the edges (and the resulting upstream QHECs) would not solely originate from an inhomogeneous screening of the back gate potential (fringing field model) but should also be due to the presence of impurities located at the sample edges (impurity model). 

The possibility of charge carriers accumulated along the edges has also been raised by Cui \textit{et al.} to explain some asymmetries between electrons and holes in their microwave impedance microscopy (MIM) data \cite{Cui2016}. In particular, they observed that $G_{xy}$ plateaus do not necessarily coincide with an insulating bulk in graphene (imaged in MIM). Depending on the device, the offset between transport and MIM signatures can be visible for holes and not for electrons, or conversely (see the supplemental Figs. S5 to S7 in \cite{Cui2016}). Nevertheless, the offset is always proportional to the back gate voltage $V_{bg}$ which indicates that the increase of charge carriers at the edges is also proportional by a factor $\eta$ to $V_{bg}$. This proportionality is coherent with the fringing field model but the electron-hole asymmetry of the parameter $\eta$ remains to be explained. For this purpose, they assume that unipolar states, associated to impurities located near the edges, hold some of the accumulated charges and decrease the accumulation profile. Their results therefore suggest an interplay between both the fringing field and the impurity models.

Based on our results, we have proposed another scenario to explain the electron-hole asymmetry observed by Cui \textit{et al.}, only considering the fringing field model (i.e. upstream QHECs exist for both charge carriers types). In the manuscript, we explain that metallic contacts induce a doping in the sample that modifies the structure of QHECs. If graphene is doped with holes below the contacts, an offset between the $G_{xy}$ plateaus and the insulating bulk is expected. Indeed, as discussed in Fig. 4 of the manuscript, the breakdown of the QHECs topological protection occurs by coupling up- and downstream modes and is independent from the bulk state. For the electrons, however, $G_{xy}$ plateaus coincide with an insulating bulk since backscattering should occur through the bulk. The inverse is expected if graphene is doped with electrons below the contacts. Our model is therefore coherent with the observations of Cui \textit{et al.}


\section{Upstream modes with contacts modeled by ideal leads}\label{ideal_leads}
\begin{figure}[!ht]
\centering
\includegraphics[width=\linewidth]{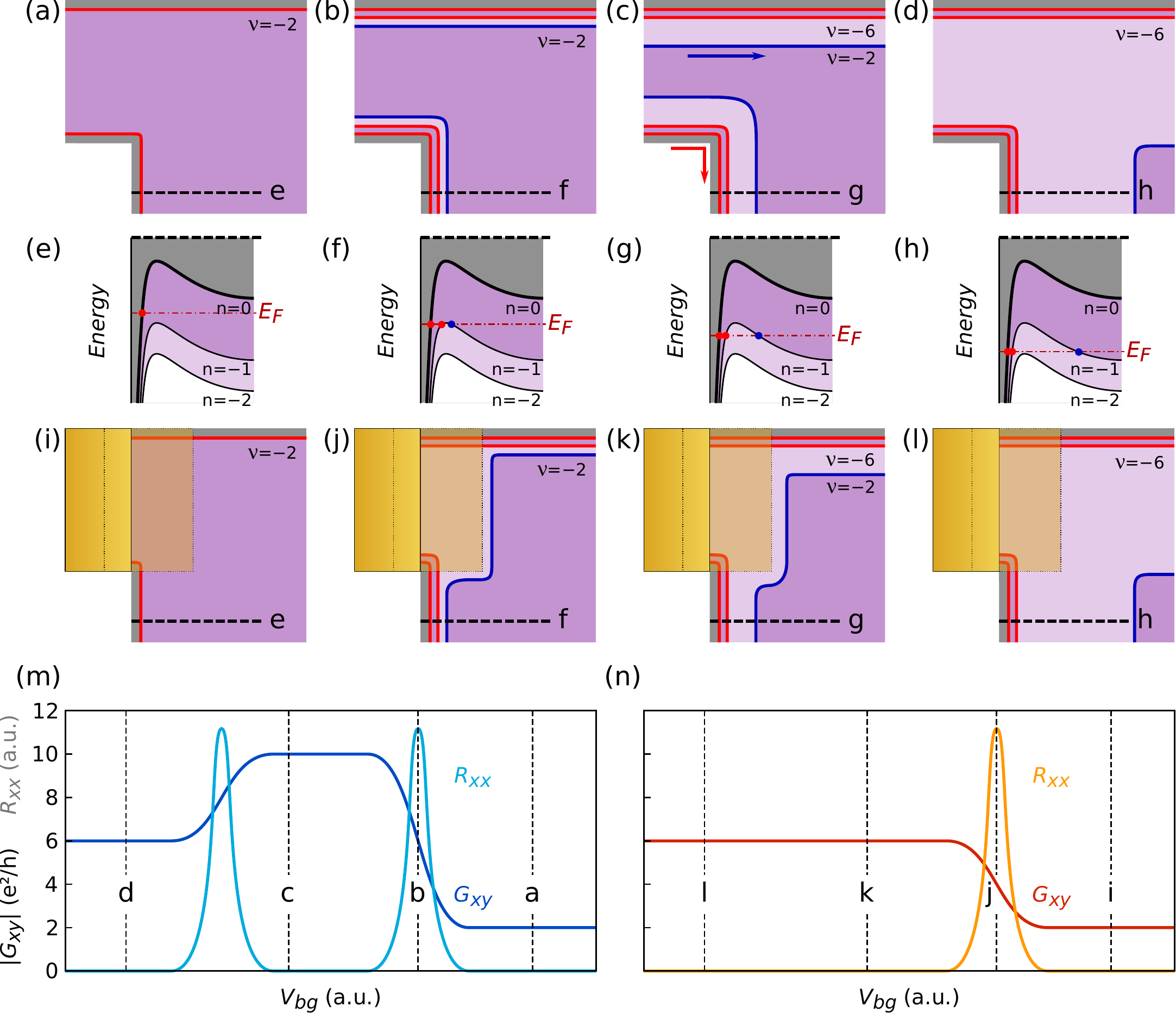}
\caption{Illustration of the QHECs spatial structure at the transition between $\nu = -2$ and $\nu = -6$ for two models of contacts. \textbf{(a-d)} Downstream (red) and upstream (blue) QHECs in the case of ideal contacts, smoothly connected with the  sample. \textbf{(e-h)} Potential profile (thick line) along the edges and the three first Landau levels ($n=0$ to $n=-2$, in thin lines). QHECs appears where the Fermi energy (red dashed line) cross the Landau levels. \textbf{(i-l)} Similar schematics as (a-d) for line contacts. Graphene is p-doped in the region located below the metallic contact. \textbf{(m,n)} Schematic evolution of longitudinal resistance $R_{xx}$ and Hall conductance $G_{xy}$ as a function of $V_{bg}$ in the case of the ideal contacts (m) and in the case of the line contacts (n).}
\label{contacts_model}
\end{figure}

As explained at the end of section \ref{imp_model}, the doping induced by contacts is a crucial ingredient to get a good picture of the QHECs spatial structure on the hole side, both for the fringing fields and for the impurity model. In this section, we compare a case where contacts are ideal (i.e., no contact-induced doped region) with the case discussed in the manuscript, where contacts dope graphene with holes. We compare both cases in Fig. \ref{contacts_model} and we show that ideal contacts should lead to peculiar features in the transport measurements, in contradiction with experimental observations. 

We start by showing in Figs. \ref{contacts_model}a-d the spatial structure of QHECs in the case of ideal contacts. The four schematics represent different configurations along the transition between $\nu = -2$ and $\nu = -6$. Schematic curves of the longitudinal resistance $R_{xx}$ and the Hall conductance $G_{xy}$ for this transition are shown in Fig. \ref{contacts_model}m. We now discuss the different configurations:
\begin{itemize}
    \item In Fig. \ref{contacts_model}a, one downstream QHEC flow along the edges, since the Fermi energy crosses only the $n=0$ Landau level (LL) in Fig. \ref{contacts_model}e. The filling factor in the insulating bulk is $\nu = -2$ and $G_{xy} = -2e^2/h$, while $R_{xx} = 0$.
    \item In Fig. \ref{contacts_model}b, both a down- and upstream QHEC appear along the edges, because the Fermi energy starts to cross the $n= -1$ LL. In this situation, the down- and upstream QHECs are close to each other so they can easily be coupled, leading to holes backscattering. $R_{xx}$ is therefore not zero as shown in Fig. \ref{contacts_model}m while $|G_{xy}|$ increases due to the emergence of new conduction channels.
    \item In Fig. \ref{contacts_model}c, the distance between the down- and upstream QHECs increases by changing the position of the Fermi energy in Fig. \ref{contacts_model}g. The backscattering between the counterpropagating QHECs is suppressed and $R_{xx}$ falls back to zero in Fig. \ref{contacts_model}m. However, $G_{xy} = -10 e^2/h$ since the downstream QHECs contribute for $-6e^2/h$ (red arrow) and the upstream QHEC contributes for $-4e^2/h$ (blue arrow). 
    \item In Fig. \ref{contacts_model}d, the upstream QHEC is even farther away from the edges (or simply disappears for a lower edge accumulation than that of Fig. \ref{contacts_model}h). So, it does not enter the contacts anymore. The filling factor of the insulating bulk is $\nu = -6$ and $G_{xy} = -6e^2/h$ since only the downstream QHECs contribute to the Hall conductance. In Fig. \ref{contacts_model}m, $R_{xx}$ therefore goes through a second peak at the transition between the two $G_{xy}$ plateaus, and then goes back to zero at more negative $V_{bg}$, when the upstream QHEC is no longer coupled to contacts. 
\end{itemize}
Obviously, the $G_{xy} = -10e^2/h$ plateau described in Fig. \ref{contacts_model}m is not observed in experiments reported in the literature. It then appears that the upstream QHEC does not act as a conduction channel between the contacts, so that it does not contribute to $G_{xy}$. This is indeed the case of the model presented in the manuscript and in Figs. \ref{contacts_model}i-l where the doping induced by the contact prevents their coupling with the upstream QHEC. The corresponding $R_{xx}$ and $G_{xy}$ curves are shown in Fig. \ref{contacts_model}n. In this situation, the transition between the $-2e^2/h$ and $-6e^2/h$, accompanied by a non zero $R_{xx}$, occurs upon the appearance of the upstream QHEC, in Fig. \ref{contacts_model}j. Indeed, the coupling between QHECs is made possible and charge carriers can be transmitted from one side of the sample to the other, as discussed in Fig. 4a,i of the manuscript. This behavior is more consistent with the experimental observations.

\section{Doping induced by the contacts}
In the manuscript, we explain that the spatial structure of QHECs can be modified due to the doping induced in graphene by the overlapping metal in the vicinity of line contacts. We have quantified this doping by computing the change of potential $\Delta \phi$ below the contacts thanks to Eq. 1 of the manuscript. This equation is based on the workfunctions of graphene and chromium determined experimentally. From this equation, we find a value $\Delta \phi = 0.007$ eV. 

However, Eq. 1 certainly fails to give a precise value of $\Delta \phi$. Indeed, the values of the workfunctions taken from the literature (especially that of chromium, less studied) depend on the experimental method. For instance, the chromium workfunction has been measured in \cite{Waldrop1984} by evaporating the metal on a clean doped GaAs surface. However, there is no guarantee that it would remain the same if measured with a graphene device. 

Performing ab-initio simulations is another way to assess chromium workfunction, as done in \cite{Giovannetti2008, Chaves2014} for other metals. For chromium, nevertheless, the task is far from being trivial. Indeed, chromium presents different crystal orientations whose workfunctions are different, as studied more extensively for other metals \cite{Gao2017,Jooya2019}. It is therefore difficult to know the relative weight of each crystalline phase in the final workfunction value.

A last consideration concerns the fact that impurities can be trapped between hBN and the metallic layer of the contacts during the fabrication process. It can also influence the accumulation of charge carriers density (either holes or electrons) below the contacts. In the framework developed in the manuscript, it can explain the asymmetry between holes and electrons observed by Cui et al. \cite{Cui2016} and discussed in section \ref{imp_model}.

\section{Tight-binding simulations}\label{suppl_simus}
As indicated in the manuscript, simulations were performed in the tight-binding framework using the KWANT package \cite{Groth2014}. The simulated system has been built from a honeycomb lattice to capture the electronic properties of graphene. In order to decrease the computation time, a scaling $s$ has been applied on the lattice parameter $a$ and the hopping parameter $t$ as
\begin{equation}\label{scaling}
    a_{\text{scaled}} = a\times s \hspace{1cm} \text{and} \hspace{1cm} t_{\text{scaled}} = t/ s
\end{equation}
These two relations keep the ratio $E_F/k_F = 3at/2$ unchanged so that graphene electronic properties are kept the same \cite{Liu2015a}. Here, the scaling factor is $s=10$.

The magnetic field $B$ has been introduced by changing the hopping parameter $t$ as
\begin{equation}
    t \longrightarrow t\exp \left( -j \frac{e}{h}B(x_i-x_j)\dfrac{y_i+y_j}{2} \right)
\end{equation}
where $x_i, y_i$ are the coordinates of the $i$th atom of the system.

The system geometry consists in a bow tie shaped geometry with a constriction width of 300 nm (see Fig. 4 of the manuscript), similar to the real device. An increase of charge carriers density has been qualitatively defined along the edges to account for the inhomogeneous screening of the back gate voltage observed experimentally. This increase is proportional to the bulk density by a factor\footnote{The local charge carrier density $n_i$ is proportional to the square of the onsite parameter $U_i^2$ where $i$ is the atoms label. An increase of charge carriers density of 0.69 is defined by an increase of the onsite parameter of 0.3 compared to that of the bulk.} $\eta = 0.69$ and has a width, take from the edges, of about 75 nm, as depicted in Figs. \ref{LDOS_up}b,c. 

Three antidots have been introduced with variations of the onsite potential following a Gaussian function, as illustrated in Figs. 4e,f of the manuscript. The radius of these Gaussian functions is 40 nm and their amplitude is -0.02 eV, so that all induce an decrease of holes density (or increase of electrons density). Two antidots are located along the edges, to induce the backscattering discussed in Fig. 4i of the manuscript. The third one is located at the middle of the constriction, to induce the backscattering shown in Fig. 4j of the manuscript.

In Figs. 4e-j of the manuscript, the \textit{measurement} leads used to compute $R_{xx}$ and $G_{xy}$ are much thinner than the \textit{injection} leads located at the left and right bottom corners of the system, used to inject and collect current (modeled by plane waves in Kwant). The reason is that one must prevent current to leave the sample through the measurement leads when computing the local current density maps presented in Figs. 4g-j of the manuscript, to correctly capture the trajectories of charge carriers. These leads are therefore chosen thin enough to prevent plane waves to enter or leave them. However, the width of the measurement leads does not affect the calculation of $R_{xx}$ (Fig. 5a of the manuscript). Indeed, it is possible to compute $R_{xx}$ by imposing a zero current in the three measurement leads, while all the current enters and leaves the system by the two injection leads. We have confirmed this by computing $R_{xx}$ for different widths of the measurement leads.

\bibliographystyle{naturemag}